\documentclass[12pt]{article}
\usepackage{amsmath}
\usepackage{amstext}
\usepackage{amsfonts}
\usepackage{amsbsy}
\usepackage{eucal}
\usepackage{amssymb}
\usepackage{amsthm}
\usepackage{color}
\usepackage{graphics}

\theoremstyle{definition}

\begin{document}
\begin{center}
{\Large  Coupled Ablowitz-Ladik equations with branched dispersion}
\end{center}
\begin{center}
Corina N. Babalic$^{*_, **}$, A. S. Carstea${^*}$
\end{center}
\begin{center}
{\it * Dept. of Theoretical Physics, Institute of Physics and Nuclear Engineering, P.O. BOX MG-6, Magurele, Bucharest, Romania\\
** University of Craiova, 13 A.I. Cuza, 200585, Craiova, Romania }
\end{center}

\begin{abstract}
Complete integrability and multisoliton solutions are discussed for a  multicomponent Ablowitz-Ladik system with branched dispersion relation. 
It is also shown that starting from a  ``diagonal'' (in two-dimensions) completely integrable Ablowitz-Ladik equation, one can obtain all the results using a periodic reduction.
\end{abstract}

\section{Introduction}
The study of discrete nonlinear Schr\"odinger equation is extremely important since the seventies. The equation appeared for the first time in the biophysical context \cite{davy} and 
since then the topic developed rapidly in nonlinear optics, Bose-Einstein condensates, etc. Initially, it was used for dynamics of wave guide arrays \cite{chr} and then in the study of 
discrete diffraction, Peierls barriers, dispersion management, etc. \cite{eis}. It was shown afterwards that nonlinear localized waves in optically-induced lattices and photo refractive
media are accurately described by discrete nonlinear Schr\"odinger equation (NLS).
However, these phenomena are more or less described by the non-integrable variant of discrete NLS. But quite recently it was shown that motion of discrete curves 
(modelling polymer's motions) can be modelled by the completely integrable variant of discrete NLS, namely Ablowitz-Ladik \cite{doliwa} and Hirota-Tsujimoto \cite{kaji} equations. 
Integrable coupled semi-discrete and discrete equations represent a topic which is still not well developed.  They have a very interesting phenomenology because of the 
extra-degrees of freedom coming from the matrix structure. Coupled soliton equations have been proposed by several authors \cite{yon,ohta,iwao,svin}. Integrable 
discretization of such systems is in general quite complicated. The Hirota bilinear formalism \cite{hirbook} turns out to be very effective inasmuch as the 
bilinear form can be discretized straightforwardly and multisoliton solution or B\"acklund transformations can be obtained directly.

In this paper we are discussing a generalisation of the Ablowitz-Ladik model \cite{AL} to the multicomponent (matrix) case, but with many branches of 
dispersion relation (being different from the well known discrete Manakov system \cite{barbara}). We will show complete integrability of this system using 
Hirota bilinear formalism and displaying the Lax pair. The main feature of such system is the structure of dispersion relation (having multiple branches) 
and phases of the components parametrised by the order $N$ roots of unity. The existence of many branches of the dispersion relation allows more freedom in 
soliton interaction. We do the analysis directly on the system using Hirota bilinear formalism and we provide also the Lax pairs proving thus the complete 
integrability. Furthermore we will show how starting from a {\it diagonal} Ablowitz-Ladik (having a diagonal one-dimensional evolution in two discrete independent 
variables) one can obtain all the results, in a very simple way, using a periodic reduction on the second discrete independent variable. The paper is organised 
as follows: in the first and second chapter we discuss the soliton solutions and Lax pairs for two and three component system and in the third one we discuss the general case and the periodic reduction.

\section{Coupled Ablowitz-Ladik equations}
The general differential-difference Ablowitz-Ladik system with branched dispersion is:

\begin{equation}\label{sys}
i\frac{d}{dt}Q_n=(1+Q_n Q_n^{*})(E_{\sigma_1}Q_{n+1}E_{\sigma_2}+E_{\sigma_2}Q_{n-1}E_{\sigma_1}),
\end{equation}
where $Q(n,t)$ is a diagonal matrix of complex functions $q_{i}(n,t)$ given by:
$$
Q_n=\left(
\begin{array}{ccccccc}
q_1(n,t)&0&0&......&0\\
0&q_2(n,t)&0&......&0\\
0&0&q_3(n,t)&......&0\\
....&....&....&.....&......&\\
0&0&0&......&q_N(n,t)\\
\end{array}\right)
$$
and $E_{\sigma_i}$  are permutation matrices corresponding to the following permutations:
$$
\sigma_1=\left(
\begin{array}{ccccccc}
1&2&.&.&.&.&N\\
2&3&.&.&.&.&1\\
\end{array}\right),\quad
\sigma_2=\left(
\begin{array}{ccccccc}
1&2&.&.&.&.&N\\
N&1&2&.&.&.&N-1\\
\end{array}\right).
$$
On the components, system (\ref{sys}) has the following expression:
\begin{eqnarray}\label{sys1}
i\dot q_1&=&(1+|q_1|^2)(\overline{q_2}+\underline{q_N})\nonumber \\
i\dot q_2&=&(1+|q_2|^2)(\overline{q_3}+\underline{q_1})\nonumber \\
&.....&\nonumber \\
i\dot q_{N-1}&=&(1+|q_{N-1}|^2)(\overline{q_N}+\underline{q_{N-2}})\\
i\dot q_N&=&(1+|q_N|^2)(\overline{q_1}+\underline{q_{N-1}})\nonumber 
\end{eqnarray}
where $q_{\mu}=q_{\mu}(n,t)$ and $\overline{q_{\mu}}=q_{\mu}(n+1,t), \underline{q_{\mu}}=q_{\mu}(n-1,t)$ for any $\mu=1,...,N$.

\vspace{0.2cm}
For $N=1$, system (\ref{sys1}) reduces to the classical Ablowitz-Ladik (AL) equation:
\begin{eqnarray}\label{1 system-NLS}
i \dot q_1&=&(1+|q_1|^2)(\overline {q_{1}}+\underline {q_{1}}),
\end{eqnarray}
for which the Lax pair and the multisoliton solutions are known  \cite{AL}. 

For $N=2$, we get the following system:
\begin{eqnarray}\label{2 system-NLS}
i \dot q_1&=&(1+|q_1|^2)(\overline {q_{2}}+\underline {q_{2}})\nonumber \\
i \dot q_2&=&(1+|q_2|^2)(\overline {q_{1}}+\underline {q_{1}})
\end{eqnarray}
which, to our knowledge, was very little investigated. The main goal of this paper is to investigate the integrability and the multisoliton solutions for the general case.
 There are many methods for investigating the integrability of a dynamical system. One of them consists in a direct computation of the conserved quantities \cite{RC1}. 
 Another one is based on the computation of Lie symmetries of the system \cite{ RC2}, \cite{ DA}. Also very interesting results 
 appeared about symmetries and conservation laws on (semi)discrete nonlinear Schr\"odinger equation in \cite{Zhang}, \cite{wei}
 
 In this paper we are going to use the Hirota bilinear formalism and we will provide the Lax pairs as well.

\subsection{Soliton solutions and Lax pairs for $N=2$}

We consider the coupled Ablowitz-Ladik with two equations (\ref{2 system-NLS}), for which we apply the Hirota bilinear formalism. 
Using the nonlinear substitutions $q_1=G_1/F_1$ and $q_2=G_2/F_2$, we cast (\ref{2 system-NLS}) into the bilinear form:
\begin{eqnarray}\label{2 system-NLS bilinear}
i {\bf D_t} G_1 \cdot F_1&=&\overline{G_2} \underline{F_2} + \underline{G_2} \overline{F_2} \nonumber\\
i {\bf D_t} G_2 \cdot F_2&=&\overline{G_1} \underline{F_1} + \underline{G_1} \overline{F_1} \nonumber\\
F_1^2+|G_1|^2&=&\overline{F_2} \underline{F_2} \nonumber\\
F_2^2+|G_2|^2&=&\overline{F_1} \underline{F_1}
\end{eqnarray}
where $F_1, F_2$ are real functions, while $G_1, G_2$ and complex valued functions. In order to solve this bilinear system we take for the one soliton solution the following ansatz: 
$$G_1=\alpha_1 e^{\eta_1}, \quad G_2=\alpha_2 e^{\eta_1}, \quad F_1=1+e^{\eta_1+\eta_1^*+\phi_1}, \quad F_2=1+e^{\eta_1+\eta_1^*+\phi_2}$$
where amplitudes are given by the components of the ``polarization vector'' $(\alpha_1,\alpha_2)$. From the first two bilinear 
equations we get an homogeneous algebraic system with the unknowns $\alpha_1,\alpha_2$. Its compatibility condition gives the dispersion relation:
$$ \omega_1=-2 i \epsilon \cosh(k_1), \quad \epsilon \in \{\pm1\}.$$ 
Next, two bilinear equations give the following two possible relations $e^{\phi_1}=e^{\phi_2}$ and $e^{\phi_1}=-e^{\phi_2}$. 
The last one gives $\alpha_1=\alpha_2=0$ so remains only $\phi_1=\phi_2\equiv \phi_{12}$. Also, from the first bilinear equations we get
$\epsilon\alpha_1=\alpha_2$. Now fixing $\alpha_1=1$ we can write down the one soliton solution in the simplest form:
\begin{eqnarray}
G_1=e^{\eta_1}, \quad G_2=\epsilon e^{\eta_1}, \quad F_1=F_2=1+e^{\eta_1+\eta_1^*+\phi_{12}} \nonumber
\end{eqnarray}
where:
$$e^{\phi_{12}}=\frac{1}{2[\cosh(k_1+k_1^*)-1]}$$
and
$$\eta_1=k_1 n+ \omega_1 t+\eta_1^{(0)}.$$
Explicitly, the 1-soliton solution of (\ref{2 system-NLS}) is:
\begin{eqnarray}
q_1=\frac{e^{k_1 n-2 i \epsilon \cosh(k_1) t+\eta_1^{(0)}}}{1+\frac{1}{2(\cosh(k_1+k_1^*)-1)} e^ {(k_1+k_1^*) n -2 i  \epsilon [\cosh(k_1)-\cosh(k_1^*)] t+\eta_1^{(0)}+\eta_1^{(0)*}}} \nonumber\\
q_2=\frac{\epsilon e^{k_1 n-2 i \epsilon \cosh(k_1) t+\eta_1^{(0)}}}{1+\frac{1}{2(\cosh(k_1+k_1^*)-1)} e^ {(k_1+k_1^*) n -2 i  \epsilon [\cosh(k_1)-\cosh(k_1^*)] t+\eta_1^{(0)}+\eta_1^{(0)*}}} \nonumber
\end{eqnarray}
{\bf Remark:} Because physically the interest is in the square modulus amplitudes, here they are fixed (unlike the case of Manakov system where the polarization vector is free) and only the propagation direction can be different.

In constructing the two-soliton solution we have to take into account that all types of solitons must be considered (in this case both directions of propagation).
Straightforward calculation gives 2-soliton solution  that describes interaction of two solitons for any propagation direction:
\begin{eqnarray*}
G_1\!\!\!&\!=\!\!\!\!&e^{\eta_1}+e^{\eta_2}+e^{\eta_1+\eta_2+\eta_2^*+\phi_{12}+\phi_{14}+\phi_{24}} +e^{\eta_1+\eta_2+\eta_1^*+\phi_{12}+\phi_{13}+\phi_{23}}\nonumber\\
G_2\!\!\!&\!=\!\!\!\!&\epsilon_1 e^{\eta_1}+\epsilon_2 e^{\eta_2}+\epsilon_1 e^{\eta_1+\eta_2+\eta_2^*+\phi_{12}+\phi_{14}+\phi_{24}} +\epsilon_2 e^{\eta_1+\eta_2+\eta_1^*+\phi_{12}+\phi_{13}+\phi_{23}}\nonumber\\
F_1\!\!\!&\!=\!\!\!\!&1+e^{\eta_1+\eta_1^*+\phi_{13}}+e^{\eta_2+\eta_2^*+\phi_{24}}+e^{\eta_1+\eta_2^*+\phi_{14}}+ e^{\eta_2+\eta_1^*+\phi_{23}}+ e^{\eta_1+\eta_1^*+\eta_2+\eta_2^*+\sum\limits^4_{1\leq i<j}\phi_{ij}} \nonumber\\
F_2\!\!\!&\!=\!\!\!\!&1+e^{\eta_1+\eta_1^*+\phi_{13}}+e^{\eta_2+\eta_2^*+\phi_{24}}+\frac{\epsilon_1}{\epsilon_2}e^{\eta_1+\eta_2^*+\phi_{14}}+\frac{\epsilon_2}{\epsilon_1} e^{\eta_2+\eta_1^*+\phi_{23}}+ e^{\eta_1+\eta_1^*+\eta_2+\eta_2^*+\sum\limits^4_{1\leq i<j}\phi_{ij}} \nonumber
\end{eqnarray*}
where:
$$e^{\phi_{ij}} = \begin{cases} \frac{1}{2} \left[\epsilon_i \epsilon_j\cosh(k_i+k_j^*)-1\right]^{-1}, & \mbox{if } i=1,2 \mbox{ and } j=3,4; \vspace{0.3cm}\\ 
2\left[\epsilon_i \epsilon_j\cosh(k_i-k_j)-1\right], & \mbox{if } i=1,2 \mbox{ and } j=1,2\\
& \mbox{or } i=3,4 \mbox{ and } j=3,4; \end{cases}$$
\begin{equation}\label{coef}
\eta_j=k_j n+ \omega_j t+\eta_j^{(0)}, \quad k_{j+2}=k_j^*
\end{equation}
$$ \omega_j=-2 i \epsilon_j \cosh(k_j), \quad \epsilon_j \in\{-1,1\}, \quad j=1,2.$$
Long but straightforward calculations show the three-soliton solution in the form:
\begin{eqnarray*}
G_1\!\!\!&\!=\!\!\!\!&e^{\eta_1}+e^{\eta_2}+e^{\eta_3}+\nonumber\\
&+& e^{\eta_1+\eta_2+\eta_2^*+\phi_{12}+\phi_{15}+\phi_{25}} +e^{\eta_2+\eta_1+\eta_1^*+\phi_{12}+\phi_{14}+\phi_{24}}+e^{\eta_3+\eta_1+\eta_1^*+\phi_{13}+\phi_{14}+\phi_{34}}+\nonumber\\
&+&e^{\eta_1+\eta_3+\eta_3^*+\phi_{13}+\phi_{16}+\phi_{36}}+e^{\eta_2+\eta_3+\eta_3^*+\phi_{23}+\phi_{26}+\phi_{36}} +e^{\eta_3+\eta_2+\eta_2^*+\phi_{23}+\phi_{25}+\phi_{35}}+\nonumber\\
&+&e^{\eta_1+\eta_2+\eta_3^*+\phi_{12}+\phi_{16}+\phi_{26}} +e^{\eta_1+\eta_3+\eta_2^*+\phi_{13}+\phi_{15}+\phi_{35}}+e^{\eta_2+\eta_3+\eta_1^*+\phi_{23}+\phi_{24}+\phi_{34}}+\nonumber\\
&+&e^{\eta_1+\eta_2+\eta_3+\eta_1^*+\eta_2^*+\phi_{12}+\phi_{13}+\phi_{14}+\phi_{15}+\phi_{23}+\phi_{24}+\phi_{25}+\phi_{34}+\phi_{35}+\phi_{45}}+ \nonumber\\
&+&e^{\eta_1+\eta_2+\eta_3+\eta_1^*+\eta_3^*+\phi_{12}+\phi_{13}+\phi_{14}+\phi_{16}+\phi_{23}+\phi_{24}+\phi_{26}+\phi_{34}+\phi_{36}+\phi_{46}}+ \nonumber\\
&+&e^{\eta_1+\eta_2+\eta_3+\eta_2^*+\eta_3^*+\phi_{12}+\phi_{13}+\phi_{15}+\phi_{16}+\phi_{23}+\phi_{25}+\phi_{26}+\phi_{35}+\phi_{36}+\phi_{56}} \nonumber
\end{eqnarray*}
\begin{eqnarray*}
G_2\!\!\!&\!=\!\!\!\!&e^{\eta_1}+e^{\eta_2}+e^{\eta_3}+\nonumber\\
&+& \epsilon_1 e^{\eta_1+\eta_2+\eta_2^*+\phi_{12}+\phi_{15}+\phi_{25}} +\epsilon_2 e^{\eta_2+\eta_1+\eta_1^*+\phi_{12}+\phi_{14}+\phi_{24}}+\epsilon_3 e^{\eta_3+\eta_1+\eta_1^*+\phi_{13}+\phi_{14}+\phi_{34}}+\nonumber\\
&+&\epsilon_1 e^{\eta_1+\eta_3+\eta_3^*+\phi_{13}+\phi_{16}+\phi_{36}}+\epsilon_2 e^{\eta_2+\eta_3+\eta_3^*+\epsilon_3 \phi_{23}+\phi_{26}+\phi_{36}} +\epsilon_3 e^{\eta_3+\eta_2+\eta_2^*+\phi_{23}+\phi_{25}+\phi_{35}}+\nonumber\\
&+&\frac{\epsilon_1 \epsilon_2}{\epsilon_3}e^{\eta_1+\eta_2+\eta_3^*+\phi_{12}+\phi_{16}+\phi_{26}} +\frac{\epsilon_1 \epsilon_3}{\epsilon_2}e^{\eta_1+\eta_3+\eta_2^*+\phi_{13}+\phi_{15}+\phi_{35}}+\frac{\epsilon_2 \epsilon_3}{\epsilon_1}e^{\eta_2+\eta_3+\eta_1^*+\phi_{23}+\phi_{24}+\phi_{34}}+\nonumber\\
&+&\epsilon_1 e^{\eta_1+\eta_2+\eta_3+\eta_2^*+\eta_3^*+\phi_{12}+\phi_{13}+\phi_{15}+\phi_{16}+\phi_{23}+\phi_{25}+\phi_{26}+\phi_{35}+\phi_{36}+\phi_{56}}+ \nonumber\\
&+&\epsilon_2 e^{\eta_1+\eta_2+\eta_3+\eta_1^*+\eta_3^*+\phi_{12}+\phi_{13}+\phi_{14}+\phi_{16}+\phi_{23}+\phi_{24}+\phi_{26}+\phi_{34}+\phi_{36}+\phi_{46}}+ \nonumber\\
&+&\epsilon_3 e^{\eta_1+\eta_2+\eta_3+\eta_1^*+\eta_2^*+\phi_{12}+\phi_{13}+\phi_{14}+\phi_{15}+\phi_{23}+\phi_{24}+\phi_{25}+\phi_{34}+\phi_{35}+\phi_{45}}\nonumber\\
\end{eqnarray*}
\begin{eqnarray*}
F_1\!\!\!&\!=\!\!\!\!&1+e^{\eta_1+\eta_1^*+\phi_{14}}+e^{\eta_2+\eta_2^*+\phi_{25}}+e^{\eta_3+\eta_3^*+\phi_{36}}+e^{\eta_1+\eta_2^*+\phi_{15}}+ e^{\eta_1+\eta_3^*+\phi_{16}}+ \nonumber\\
&+&e^{\eta_2+\eta_1^*+\phi_{24}}+e^{\eta_2+\eta_3^*+\phi_{26}}+e^{\eta_3+\eta_1^*+\phi_{34}}+e^{\eta_3+\eta_2^*+\phi_{35}}+\nonumber\\
&+&e^{\eta_1+\eta_2+\eta_1^*+\eta_2^*+\phi_{12}+\phi_{14}+\phi_{15}+\phi_{24}+\phi_{25}+\phi_{45}}+e^{\eta_1+\eta_2+\eta_1^*+\eta_3^*+\phi_{12}+\phi_{14}+\phi_{16}+\phi_{24}+\phi_{26}+\phi_{46}}+\nonumber\\
&+&e^{\eta_1+\eta_2+\eta_2^*+\eta_3^*+\phi_{12}+\phi_{15}+\phi_{16}+\phi_{25}+\phi_{26}+\phi_{56}}+e^{\eta_1+\eta_3+\eta_1^*+\eta_2^*+\phi_{13}+\phi_{14}+\phi_{15}+\phi_{34}+\phi_{35}+\phi_{45}}+\nonumber\\
&+&e^{\eta_1+\eta_3+\eta_1^*+\eta_3^*+\phi_{13}+\phi_{14}+\phi_{16}+\phi_{34}+\phi_{36}+\phi_{46}}+e^{\eta_1+\eta_3+\eta_2^*+\eta_3^*+\phi_{13}+\phi_{15}+\phi_{16}+\phi_{35}+\phi_{36}+\phi_{56}}+\nonumber\\
&+&e^{\eta_2+\eta_3+\eta_1^*+\eta_2^*+\phi_{23}+\phi_{24}+\phi_{26}+\phi_{34}+\phi_{35}+\phi_{45}}+e^{\eta_2+\eta_3+\eta_1^*+\eta_3^*+\phi_{23}+\phi_{24}+\phi_{26}+\phi_{34}+\phi_{36}+\phi_{46}}+\nonumber\\
&+&e^{\eta_2+\eta_3+\eta_2^*+\eta_3^*+\phi_{23}+\phi_{25}+\phi_{26}+\phi_{35}+\phi_{36}+\phi_{56}}+e^{\eta_1+\eta_2+\eta_3+\eta_1^*+\eta_2^*+\eta_3^*+\sum\limits^6_{1\leq i<j}\phi_{ij}} \nonumber
\end{eqnarray*}
\begin{eqnarray*}
F_2\!\!\!&\!=\!\!\!\!&1+e^{\eta_1+\eta_1^*+\phi_{14}}+e^{\eta_2+\eta_2^*+\phi_{25}}+e^{\eta_3+\eta_3^*+\phi_{36}}+\frac{\epsilon_1}{\epsilon_2}e^{\eta_1+\eta_2^*+\phi_{15}}+ \frac{\epsilon_1}{\epsilon_3}e^{\eta_1+\eta_3^*+\phi_{16}}+ \nonumber\\
&+&\frac{\epsilon_2}{\epsilon_1}e^{\eta_2+\eta_1^*+\phi_{24}}+\frac{\epsilon_2}{\epsilon_3}e^{\eta_2+\eta_3^*+\phi_{26}}+\frac{\epsilon_3}{\epsilon_1}e^{\eta_3+\eta_1^*+\phi_{34}}+\frac{\epsilon_3}{\epsilon_2}e^{\eta_3+\eta_2^*+\phi_{35}}+\nonumber\\
&+&e^{\eta_1+\eta_2+\eta_1^*+\eta_2^*+\phi_{12}+\phi_{14}+\phi_{15}+\phi_{24}+\phi_{25}+\phi_{45}}+\frac{\epsilon_2}{\epsilon_3}e^{\eta_1+\eta_2+\eta_1^*+\eta_3^*+\phi_{12}+\phi_{14}+\phi_{16}+\phi_{24}+\phi_{26}+\phi_{46}}+\nonumber\\
&+&\frac{\epsilon_1}{\epsilon_3}e^{\eta_1+\eta_2+\eta_2^*+\eta_3^*+\phi_{12}+\phi_{15}+\phi_{16}+\phi_{25}+\phi_{26}+\phi_{56}}+\frac{\epsilon_3}{\epsilon_2}e^{\eta_1+\eta_3+\eta_1^*+\eta_2^*+\phi_{13}+\phi_{14}+\phi_{15}+\phi_{34}+\phi_{35}+\phi_{45}}+\nonumber\\
&+&e^{\eta_1+\eta_3+\eta_1^*+\eta_3^*+\phi_{13}+\phi_{14}+\phi_{16}+\phi_{34}+\phi_{36}+\phi_{46}}+\frac{\epsilon_1}{\epsilon_2}e^{\eta_1+\eta_3+\eta_2^*+\eta_3^*+\phi_{13}+\phi_{15}+\phi_{16}+\phi_{35}+\phi_{36}+\phi_{56}}+\nonumber\\
&+&\frac{\epsilon_3}{\epsilon_1}e^{\eta_2+\eta_3+\eta_1^*+\eta_2^*+\phi_{23}+\phi_{24}+\phi_{26}+\phi_{34}+\phi_{35}+\phi_{45}}+\frac{\epsilon_2}{\epsilon_1}e^{\eta_2+\eta_3+\eta_1^*+\eta_3^*+\phi_{23}+\phi_{24}+\phi_{26}+\phi_{34}+\phi_{36}+\phi_{46}}+\nonumber\\
&+&e^{\eta_2+\eta_3+\eta_2^*+\eta_3^*+\phi_{23}+\phi_{25}+\phi_{26}+\phi_{35}+\phi_{36}+\phi_{56}}+e^{\eta_1+\eta_2+\eta_3+\eta_1^*+\eta_2^*+\eta_3^*+\sum\limits^6_{1\leq i<j}\phi_{ij}} \nonumber
\end{eqnarray*}
where:
$$e^{\phi_{ij}} = \begin{cases} \frac{1}{2} \left[\epsilon_i \epsilon_j\cosh(k_i+k_j^*)-1\right]^{-1}, & \mbox{if } i=1,2,3 \mbox{ and } j=4,5,6; \vspace{0.3cm}\\ 
2\left[\epsilon_i \epsilon_j\cosh(k_i-k_j)-1\right], & \mbox{if } i=1,2,3 \mbox{ and } j=1,2,3\\
& \mbox{or } i=4,5,6 \mbox{ and } j=4,5,6; \end{cases}$$
\begin{equation}\label{coef}
\eta_j=k_j n+ \omega_j t+\eta_j^{(0)}, \quad k_{j+3}=k_j^*
\end{equation}
$$ \omega_j=-2 i \epsilon_j \cosh(k_j), \quad \epsilon_j \in\{-1,1\}, \quad j=1,2,3.$$
\vspace{0.5cm}

This solution describes interaction of three solitons for any propagation direction. 
Although the existence of three-soliton solution in Hirota form for NLS-type equations is a strong indicator for the complete integrability \cite{jarmo}, we also give here the Lax pair:

\begin{eqnarray}\label{Lax 2 cuplate}
L_{n}=\left(
\begin{array}{cccc}
0&z&0&q_2\\
z&0&q_1&0\\
0&-q_2^*&0&z^{-1}\\
-q_1^*&0&z^{-1}&0\\
\end{array}\right),
\end{eqnarray}

$$B_n=\left(
\begin{array}{cccccc}
\!\!\!i(Z-q_1\underline{q_2^*})\!\!&\!\!0\!\!&\!\!i(-zq_1+z^{-1}\underline{q_2})\!\!&\!\!0\!\!\\
\!\!\!0\!&\!\!i(Z-q_2\underline{q_1^*})\!\!&\!\!0\!\!&\!\!i(-zq_2+z^{-1}\underline{q_1})\!\!\\
\!\!\!i(z\underline{q^*_2}-z^{-1}q_1^*)\!\!&\!\!0\!\!&\!\!-i(Z-q_1^*\underline{q_2})\!\!&\!\!0\!\!\\
\!\!\!0&\!\!i(z\underline{q^*_1}-z^{-1}q_2^*)\!\!&\!\!0\!\!&\!\!-i(Z-q_2^*\underline{q_1})\!\!\\
\end{array}\right),$$
where  $z$ is the spectral parameter and $Z=-\frac{1}{2}\left(z^2+z^{-2}\right)$. The compatibility condition  $\partial_t L+LB-\overline{B}L=0$ gives precisely system  (\ref{2 system-NLS}).

\vspace{0.5cm}
{\bf Remark:} The above spectral Lax operator can be put in a block Ablowitz-Ladik spectral operator as in {\cite{16a}} 

\vspace{0.5cm}

Because the system is completely integrable we can write easily the $N$-soliton solution:

\begin{eqnarray*}
G_1&=&\sum\limits_{\mu=0,1}D_2(\underline{\mu})\exp\left({\sum\limits_{i=1}^{2N}\mu_i \eta_i+\sum\limits^{2N}_{1\leq i<j}\mu_i\mu_j\phi_{ij}}\right)\\
G_2&=&\sum\limits_{\mu=0,1}D_2(\underline{\mu})\exp\left({\sum\limits_{i=1}^{2N}\mu_i [\eta_i+\log(\epsilon_i)]+\sum\limits^{2N}_{1\leq i<j}\mu_i\mu_j\phi_{ij}}\right)\\
F_1&=&\sum\limits_{\mu=0,1}D_1(\underline{\mu})\exp\left({\sum\limits_{i=1}^{2N}\mu_i \eta_i+\sum\limits^{2N}_{1\leq i<j}\mu_i\mu_j\phi_{ij}}\right)\\
F_2&=&\sum\limits_{\mu=0,1}D_1(\underline{\mu})\exp\left({\sum\limits_{i=1}^{2N}\mu_i [\eta_i+\log(\epsilon_i)]+\sum\limits^{2N}_{1\leq i<j}\mu_i\mu_j\phi_{ij}}\right)\\
\end{eqnarray*}
where:
$$e^{\phi_{ij}} = \begin{cases} \frac{1}{2} \left[\epsilon_i \epsilon_j\cosh(k_i+k_j^*)-1\right]^{-1}, & \mbox{if } i=1,...N \mbox{ and } j=N+1,...,2N; \vspace{0.3cm}\\ 
2\left[\epsilon_i \epsilon_j\cosh(k_i-k_j)-1\right], & \mbox{if } i=1,...N  \mbox{ and } i=1,...,N \\
& \mbox{or } i=N+1,...,2N \mbox{ and } j=N+1,...,2N; \end{cases}$$
\begin{eqnarray*}
\eta_j=k_j n+ \omega_j t+\eta_j^{(0)}, \quad \eta_{j+N}=\eta_j^*, \quad k_{j+N}=k_j^*,\\
\omega_j=-2 i \epsilon_j \cosh(k_j), \quad \omega_{j+N}=\omega_j^*,  \quad j=1,...,N\\
\log(\epsilon_{j+N})=\log(\epsilon_{j})^*, \quad \log(\epsilon_j)\in\{\pm \pi i\},
\end{eqnarray*}
and where:
$$D_1(\underline{\mu}) = \begin{cases} 1, & \mbox{when } \sum\limits_{i=1}^N \mu_i=\sum\limits_{i=1}^N \mu_{i+N}; \\
0 & \mbox{otherwise}; \end{cases}$$
$$D_2(\underline{\mu}) = \begin{cases} 1, & \mbox{when } \sum\limits_{i=1}^N \mu_i=1+\sum\limits_{i=1}^N \mu_{i+N}; \\
0 & \mbox{otherwise}. \end{cases}$$

\subsection{Soliton solutions and Lax pairs for $N=3$}

We consider the coupled Ablowitz-Ladik with three equations:
\begin{eqnarray}\label{3 system-NLS}
i \dot q_1&=&(1+|q_1|^2)(\overline {q_{2}}+\underline {q_{3}})\nonumber \\
i \dot q_2&=&(1+|q_2|^2)(\overline {q_{3}}+\underline {q_{1}})\nonumber \\
i \dot q_3&=&(1+|q_3|^2)(\overline {q_{1}}+\underline {q_{2}})
\end{eqnarray}
The system is again a completely integrable one. The Lax pair is the following

\begin{eqnarray}\label{Lax 3 cuplat}
L_n=\left(
\begin{array}{cccccc}
0&0&z&0&0&q_3\\
z&0&0&q_1&0&0\\
0&z&0&0&q_2&0\\
0&0&-q_3^*&0&0&z^{-1}\\
-q_1^*&0&0&z^{-1}&0&0\\
0&-q_2^*&0&0&z^{-1}&0\\
\end{array}\right), \quad B_n=
\end{eqnarray}
$$
\left(
\begin{array}{cccccc}
\!\!\!\!i(Z_+\!-\!q_1\underline{q_3^*})\!\!&\!\!0\!\!&\!\!0\!\!&\!\!i(-zq_1\!+\!\frac{\underline{q_3}}{z})\!\!&\!\!0\!\!&\!\!0\!\!\!\!\\
\!\!\!\!0\!\!&\!\!i(Z_+\!-\!q_2\underline{q_1^*})\!\!&\!\!0\!\!&\!\!0&\!\!i(-zq_2\!+\!\frac{\underline{q_1}}{z})\!\!&\!\!0\!\!\!\!\\
\!\!\!\!0\!\!&\!\!0\!&\!\!i(Z_+\!-\!q_3\underline{q_2^*})\!\!&\!\!0\!\!&\!\!0\!\!&\!\!i(-zq_3\!+\!\frac{\underline{q_2}}{z})\!\!\!\!\\
\!\!\!\!i(z\underline{q^*_3}\!-\!\frac{q_1^*}{z})\!\!&\!\!0\!\!&\!\!0\!\!&\!\!-i(Z_-\!-\!q_1^*\underline{q_3})\!\!&\!\!0\!&\!0\!\!\!\!\\
\!\!\!\!0\!\!&\!\!i(z\underline{q^*_1}\!-\!\frac{q_2^*}{z})\!\!&\!\!0\!\!&\!\!0\!&\!\!-i(Z_-\!-\!q_2^*\underline{q_1})\!\!&\!\!0\!\!\!\!\\
\!\!\!\!0\!\!&\!\!0\!\!&\!i(z\underline{q^*_2}\!-\!\frac{q_3^*}{z})\!\!&\!0\!\!\!&\!\!0\!&\!-i(Z_-\!-\!q_3^*\underline{q_2})\!\!\!\!\\
\end{array}\right)
$$
where $Z_+=1-z^2, Z_-=-1-z^{-2}$.

The compatibility condition $\partial_t L+LB-\overline{B}L=0$ gives precisely our system (\ref{3 system-NLS}). In the last section we will show how these Lax pairs were found.

In order to see the solutions we build the Hirota bilinear form, considering the nonlinear substitutions: $q_1=G_1/F_1$, $q_2=G_2/F_2$ and $q_3=G_3/F_3$:
\begin{eqnarray*}\label{3 system-NLS bilinear}
i {\bf D_t} G_1 \cdot F_1&=&\overline{G_2} \underline{F_3} + \underline{G_3} \overline{F_2} \nonumber\\
i {\bf D_t} G_2 \cdot F_2&=&\overline{G_3} \underline{F_1} + \underline{G_1} \overline{F_3} \nonumber\\
i {\bf D_t} G_3 \cdot F_3&=&\overline{G_1} \underline{F_2} + \underline{G_2} \overline{F_3} \nonumber\\
F_1^2+|G_1|^2&=&\overline{F_2} \underline{F_3} \nonumber\\
F_2^2+|G_2|^2&=&\overline{F_3} \underline{F_1} \nonumber\\
F_3^2+|G_3|^2&=&\overline{F_1} \underline{F_2} 
\end{eqnarray*}
where $F_1, F_2, F_3$ are real functions, while $G_1, G_2, G_3$ and complex functions.
Doing the same machinery as in the case of $N=2$, we can start with the following ansatz $G_i=\alpha_i e^{\eta}, \quad F_i=1+e^{\eta_i+\eta_i^*+\phi_i}$ for $i=1,2,3$. Plugging into 
the bilinear equations, we will find that $\phi_1=\phi_2=\phi_3\equiv \phi_{12}$ and $\alpha_1,\alpha_2,\alpha_3$ are ordered as the cubic roots of unity. Accordingly, the 1-soliton solution is:
\begin{eqnarray*}
G_1=e^{\eta_1},\quad G_2=\epsilon e^{\eta_1}, \quad G_3=\epsilon^2 e^{\eta_1}, \quad F_1=F_2=F_3=1+ e^{\eta_1+\eta_1^*+\phi_{12}}
\end{eqnarray*}
where:
$$\epsilon \in \left \{1,-\frac{1}{2}\pm i \frac{\sqrt3}{2}\right \}, \quad e^{\phi_{12}}=\frac{1}{2[\cosh(k_1+k_1^*)-1]}$$
and
$$\eta_1=k_1 n+ \omega_1 t+\eta_1^{(0)}$$
with three branches of dispersion:
\begin{eqnarray}  
\omega_1=-2 i \cosh(k_1) \hspace{1.9cm} & \mbox{for} & \epsilon=1 \nonumber \\
\omega_1= i \cosh(k_1)+\sqrt3 \sinh(k_1) & \mbox{for} & \epsilon = -\frac{1}{2}+i \frac{\sqrt 3}{2} \nonumber \\
\omega_1= i \cosh(k_1)-\sqrt3 \sinh(k_1) & \mbox{for} & \epsilon = -\frac{1}{2}-i \frac{\sqrt 3}{2} \nonumber 
\end{eqnarray}
\vspace{0.1cm}

The 2-soliton solution has the following form:
\begin{eqnarray*}
G_1\!\!\!&\!=\!\!\!\!&e^{\eta_1}+e^{\eta_2}+e^{\eta_1+\eta_2+\eta_2^*+\phi_{12}+\phi_{14}+\phi_{24}} +e^{\eta_1+\eta_2+\eta_1^*+\phi_{12}+\phi_{13}+\phi_{23}}\nonumber\\
G_2\!\!\!&\!=\!\!\!\!&\epsilon_1 e^{\eta_1}+\epsilon_2 e^{\eta_2}+\epsilon_1 e^{\eta_1+\eta_2+\eta_2^*+\phi_{12}+\phi_{14}+\phi_{24}} +\epsilon_2 e^{\eta_1+\eta_2+\eta_1^*+\phi_{12}+\phi_{13}+\phi_{23}}\nonumber\\
G_3\!\!\!&\!=\!\!\!\!&\epsilon_1^2 e^{\eta_1}+\epsilon_2^2 e^{\eta_2}+\epsilon_1^2 e^{\eta_1+\eta_2+\eta_2^*+\phi_{12}+\phi_{14}+\phi_{24}} +\epsilon_2^2 e^{\eta_1+\eta_2+\eta_1^*+\phi_{12}+\phi_{13}+\phi_{23}}\nonumber\\
F_1\!\!\!&\!=\!\!\!\!&1+e^{\eta_1+\eta_1^*+\phi_{13}}+e^{\eta_2+\eta_2^*+\phi_{24}}+e^{\eta_1+\eta_2^*+\phi_{14}}+ e^{\eta_2+\eta_1^*+\phi_{23}}+ e^{\eta_1+\eta_1^*+\eta_2+\eta_2^*+\sum\limits^4_{1\leq i<j}\phi_{ij}} \nonumber\\
F_2\!\!\!&\!=\!\!\!\!&1+e^{\eta_1+\eta_1^*+\phi_{13}}+e^{\eta_2+\eta_2^*+\phi_{24}}+\frac{\epsilon_1}{\epsilon_2}e^{\eta_1+\eta_2^*+\phi_{14}}+\frac{\epsilon_2}{\epsilon_1} e^{\eta_2+\eta_1^*+\phi_{23}}+ e^{\eta_1+\eta_1^*+\eta_2+\eta_2^*+\sum\limits^4_{1\leq i<j}\phi_{ij}} \nonumber\\
F_3\!\!\!&\!=\!\!\!\!&1+e^{\eta_1+\eta_1^*+\phi_{13}}+e^{\eta_2+\eta_2^*+\phi_{24}}+\left(\frac{\epsilon_1}{\epsilon_2}\right)^2e^{\eta_1+\eta_2^*+\phi_{14}}+\left(\frac{\epsilon_2}{\epsilon_1}\right)^2 e^{\eta_2+\eta_1^*+\phi_{23}}+\nonumber\\
 &&+e^{\eta_1+\eta_1^*+\eta_2+\eta_2^*+\sum\limits^4_{1\leq i<j}\phi_{ij}} \nonumber
\end{eqnarray*}
where:
$$e^{\phi_{ij}} = \begin{cases} \frac{1}{2} \left(\frac{\epsilon_i^2+\epsilon_j^2}{2\epsilon_i\epsilon_j}\cosh(k_i+k_j^*)+\frac{\epsilon_i^2-\epsilon_j^2}{2\epsilon_i\epsilon_j}\sinh(k_i+k_j^*)-1\right)^{-1}, \mbox{if } i=1,2 \mbox{ and } j=3,4; \vspace{0.5cm}\\ 
2\left(\frac{\epsilon_i^2+\epsilon_j^2}{2\epsilon_i\epsilon_j}\cosh(k_i-k_j)+\frac{\epsilon_i^2-\epsilon_j^2}{2\epsilon_i\epsilon_j}\sinh(k_i-k_j)-1\right),\mbox{if } i=1,2  \mbox{ and } j=1,2 \vspace{0.2cm}\\
\hspace{8.7cm} \mbox{or } i=3,4 \mbox{ and } j=3,4; \end{cases}$$
and:
\begin{eqnarray*}
&\eta_j=k_j n+ \omega_j t+\eta_j^{(0)}, \quad k_{j+2}=k_j^*, \quad j=1,2\\
&\log(\epsilon_{j+2})=\log(\epsilon_{j})^*, \quad \log(\epsilon_j)\in\{\frac{2 \pi}{3}i, \frac{4 \pi}{3}i, 2\pi i \},
\end{eqnarray*}
with the three branches of dispersion ($j=1,2$):
 \begin{eqnarray}  
\omega_j=-2 i \cosh(k_j) \hspace{1.9cm} & \mbox{for} & \epsilon_j=1 \nonumber \\
\omega_j= i \cosh(k_j)+\sqrt3 \sinh(k_j) & \mbox{for} & \epsilon_j = -\frac{1}{2}+i \frac{\sqrt 3}{2} \nonumber \\
\omega_j= i \cosh(k_j)-\sqrt3 \sinh(k_j) & \mbox{for} & \epsilon_j = -\frac{1}{2}-i \frac{\sqrt 3}{2}. \nonumber 
\end{eqnarray}
$3$-soliton solution can be constructed easily and this validates the Lax integrability.
\begin{eqnarray*}\label{3ss-3}
G_1&=&\sum\limits_{\mu=0,1}D_2(\underline{\mu})\exp\left({\sum\limits_{i=1}^{6}\mu_i \eta_i+\sum\limits^{6}_{1\leq i<j}\mu_i\mu_j\phi_{ij}}\right)\\
G_2&=&\sum\limits_{\mu=0,1}D_2(\underline{\mu})\exp\left({\sum\limits_{i=1}^{6}\mu_i [\eta_i+\log(\epsilon_i)]+\sum\limits^{6}_{1\leq i<j}\mu_i\mu_j\phi_{ij}}\right)\\
G_3&=&\sum\limits_{\mu=0,1}D_2(\underline{\mu})\exp\left({\sum\limits_{i=1}^{6}\mu_i [\eta_i+2\log(\epsilon_i)]+\sum\limits^{6}_{1\leq i<j}\mu_i\mu_j\phi_{ij}}\right)\\
F_1&=&\sum\limits_{\mu=0,1}D_1(\underline{\mu})\exp\left({\sum\limits_{i=1}^{6}\mu_i \eta_i+\sum\limits^{6}_{1\leq i<j}\mu_i\mu_j\phi_{ij}}\right)\\
F_2&=&\sum\limits_{\mu=0,1}D_1(\underline{\mu})\exp\left({\sum\limits_{i=1}^{6}\mu_i [\eta_i+\log(\epsilon_i)]+\sum\limits^{6}_{1\leq i<j}\mu_i\mu_j\phi_{ij}}\right)\\
F_3&=&\sum\limits_{\mu=0,1}D_1(\underline{\mu})\exp\left({\sum\limits_{i=1}^{6}\mu_i [\eta_i+2\log(\epsilon_i)]+\sum\limits^{6}_{1\leq i<j}\mu_i\mu_j\phi_{ij}}\right)\\
\end{eqnarray*}
where:
$$e^{\phi_{ij}} = \begin{cases} \frac{1}{2} \left(\frac{\epsilon_i^2+\epsilon_j^2}{2\epsilon_i\epsilon_j}\cosh(k_i+k_j^*)+\frac{\epsilon_i^2-\epsilon_j^2}{2\epsilon_i\epsilon_j}\sinh(k_i+k_j^*)-1\right)^{-1}\vspace{0.2cm}\\
\hspace{3cm} \mbox{if } i=1,2,3 \mbox{ and } j=4,5,6; \vspace{0.5cm}\\ 
2\left(\frac{\epsilon_i^2+\epsilon_j^2}{2\epsilon_i\epsilon_j}\cosh(k_i-k_j)+\frac{\epsilon_i^2-\epsilon_j^2}{2\epsilon_i\epsilon_j}\sinh(k_i-k_j)-1\right),\vspace{0.2cm}\\
\hspace{3cm} \mbox{if } i=1,2,3  \mbox{ and } i=1,2,3 \vspace{0.2cm}\\
\hspace{3cm} \mbox{or } i=4,5,6 \mbox{ and } j=4,5,6; \end{cases}$$
and:
\begin{eqnarray*}
\eta_j=k_j n+ \omega_j t+\eta_j^{(0)}, \quad \eta_{j+3}=\eta_j^*, \quad k_{j+3}=k_j^*, \quad j=1,...,3\\
\log(\epsilon_{j+3})=\log(\epsilon_{j})^*, \quad \omega_{j+3}=\omega_j^*, \quad \log(\epsilon_j)\in\{\frac{2 \pi}{3}i, \frac{4 \pi}{3}i, 2\pi i \}.
\end{eqnarray*}
\begin{eqnarray}\label{coeficienti generali}
D_1(\underline{\mu}) = \begin{cases} 1, & \mbox{when } \sum\limits_{i=1}^3 \mu_i=\sum\limits_{i=1}^3 \mu_{i+3}; \\
0 & \mbox{otherwise}; \end{cases}\nonumber\\
D_2(\underline{\mu}) = \begin{cases} 1, & \mbox{when } \sum\limits_{i=1}^3 \mu_i=1+\sum\limits_{i=1}^3 \mu_{i+3}; \\
0 & \mbox{otherwise}. \end{cases}
\end{eqnarray}
Since $\epsilon_j \in \left\{ 1,  -\frac{1}{2}\pm i \frac{\sqrt 3}{2} \right\}$, each of the three solitons can have any of the following three branches of dispersion: 
\begin{eqnarray}  
\omega_j=-2 i \cosh(k_j) \hspace{1.9cm} & \mbox{for} & \epsilon_j=1 \nonumber \\
\omega_j= i \cosh(k_j)+\sqrt3 \sinh(k_j) & \mbox{for} & \epsilon_j = -\frac{1}{2}+i \frac{\sqrt 3}{2} \nonumber \\
\omega_j= i \cosh(k_j)-\sqrt3 \sinh(k_j) & \mbox{for} & \epsilon_j = -\frac{1}{2}-i \frac{\sqrt 3}{2} \nonumber 
\end{eqnarray}

Extension to $N$-soliton solution is straightforward. Formulas are the same except that the sums in (\ref{3ss-3}) are taken from 1 to $2N$ and all the definitions with $i+3$ and $j+3$  will turn into $i+N$ and $j+N$

\section{General case and periodic reduction}

Starting from the general system of coupled Ablowitz-Ladik with $N$ coupled equations, given in (\ref{sys1}):
\begin{eqnarray}
i \dot q_1&=&(1+|q_1|^2)(\overline {q_{2}}+\underline {q_{N}})\nonumber \\
i \dot q_2&=&(1+|q_2|^2)(\overline {q_{3}}+\underline {q_{1}})\nonumber \\
i \dot q_3&=&(1+|q_3|^2)(\overline {q_{4}}+\underline {q_{2}}) \nonumber \\
.....&....&.......................\nonumber \\
i \dot q_{N-1}&=&(1+|q_{N-1}|^2)(\overline {q_{N}}+\underline {q_{N-2}})\nonumber \\
i \dot q_N&=&(1+|q_N|^2)(\overline {q_{1}}+\underline {q_{N-1}})\nonumber
\end{eqnarray}
and using  the nonlinear substitutions: $q_{\nu}=G_{\nu}/F_{\nu}$, ${\nu}=1,...,N$, we obtain the Hirota bilinear form:
\begin{eqnarray}\label{N-bil}
i {\bf D_t} G_{\nu} \cdot F_{\nu}&=&\overline{G_{\nu+1}} \underline{F_{\nu-1}} + \underline{G_{\nu-1}} \overline{F_{\nu+1}} \nonumber\\
F_{\nu}^2+|G_{\nu}|^2&=&\overline{F_{\nu+1}} \underline{F_{\nu-1}}. 
\end{eqnarray}
The 3-soliton solution for (\ref{sys1}) has the expressions for $G_{\nu}$, $F_{\nu}$, $(\nu=1,...,N)$:
\begin{eqnarray}\label{solutie generalaa}
G_{\nu}&=&\sum\limits_{\mu=0,1}D_2(\underline{\mu})\exp\left({\sum\limits_{i=1}^{6}\mu_i [\eta_i+(\nu-1)\log(\epsilon_i)]+\sum\limits^{6}_{1\leq i<j}\mu_i\mu_j\phi_{ij}}\right)\nonumber\\
F_{\nu}&=&\sum\limits_{\mu=0,1}D_1(\underline{\mu})\exp\left({\sum\limits_{i=1}^{6}\mu_i [\eta_i+(\nu-1)\log(\epsilon_i)]+\sum\limits^{6}_{1\leq i<j}\mu_i\mu_j\phi_{ij}}\right)
\end{eqnarray}
where $D_1(\underline{\mu})$, $D_2(\underline{\mu})$, $\eta_i$, $\phi_{ij}$ are given in (\ref{coeficienti generali})
 with the difference that now we have $N$ branches of dispersion for each soliton ($k_j$ is the wave number of the $j$-soliton where $j=1,2,3$): 
\begin{equation}  
\omega_j(k_{j})\!\!=\!\!-2 i \left[\frac{\epsilon_j^2+1}{2\epsilon_j}\cosh k_{j}+\frac{\epsilon_j^2-1}{2\epsilon_j}\sinh k_{j}\right]\!\!, \: \epsilon_j \in \left\{\!e^{l \frac{2 \pi i}{N}\!} \right\}\!,l=1,..N,  
j=1, 2, 3. \nonumber 
\end{equation}
So the branches of dispersion are labelled by the index $l$. So the parameter $\epsilon_j$ which characterizes the $j$-soliton ($j=1,2,3$) can have $N$ values, the order $N$ roots of unity. 

\subsection{The periodic reduction}
One could obtain the above results starting from a general ``diagonal'' equation in two dimensions and performing periodic reduction. 
This idea has been used for the first time in \cite{toki-fane} where coupled semidiscrete KdV equations were obtained modelling a modular genetic network. 
The idea is to consider that the independent discrete variable of ordinary Ablowitz-Ladik equation is in fact a {\it diagonal} in a two-dimensional (or $d$-dimensional) lattice. 
Imposing periodic reduction on the one such coordinate in that 2D-lattice, then we will obtain {\it coupled systems} of Ablowitz-Ladik equations. In order to see it clearly let us start with the following equation:
\begin{eqnarray}\label{general system-NLS}
i \frac{d}{dt} Q_{n,m}(t)&=&(1+|Q_{n,m}(t)|^2)(Q_{n+1,m+1}(t)+Q_{n-1,m-1}(t)).
\end{eqnarray}
This equation is of course completely integrable. It has the following Lax pair:
\begin{eqnarray}\label{Lax general}
L_{n,m}=\left(
\begin{array}{cc}
z&Q_{n,m}(t)\\
-Q^*_{n,m}(t)&1/z\\
\end{array}\right),
\end{eqnarray}
$$
B_{n,m}=\left(
\begin{array}{cc}
i(Z-Q_{n,m}(t)Q^*_{n-1,m-1}(t))&i(-zQ_{n,m}(t)+z^{-1}Q_{n-1,m-1}(t))\\
i(-z^{-1}Q^*_{n,m}(t)+zQ^*_{n-1,m-1}(t))&-i(Z-Q^*_{n,m}(t)Q_{n-1,m-1}(t))
\end{array}\right),
$$
where  $Z=-\frac{1}{2}\left(z^2+z^{-2}\right)$, the compatibility condition $\partial_t L_{n,m}+L_{n,m}B_{n,m}-B_{n+1,m+1}L_{n,m}=0$ gives exactly (\ref{general system-NLS}).

Now lets consider the periodic 2-reduction on the $m$ direction (meaning that $Q(n,m)$ is a periodic function only with respect to $m$ and the period is 2). This means 
that $Q(n,m)\equiv q_1(n), \: Q(n,m+1)\equiv q_2(n), \: Q(n,m+2)\equiv q_1(n), \: Q(n,m-1)\equiv q_2(n), \: etc.$ Introducing this reduction in (\ref{general system-NLS}) we get precisely: 
\begin{eqnarray*}
i \dot q_1&=&(1+|q_1|^2)(\overline {q_{2}}+\underline {q_{2}})\nonumber \\
i \dot q_2&=&(1+|q_2|^2)(\overline {q_{1}}+\underline {q_{1}}).
\end{eqnarray*}

In the same way, if we impose periodic-3 reduction, we get the system with three equations (\ref{3 system-NLS}) and so on. 
Accordingly, the general system can be obtained from this diagonal Ablowitz-Ladik equation (\ref{general system-NLS}), but choosing a periodic $N$-reduction on $m$.

The Hirota bilinear form can be obtained by $Q_{n,m}(t)=G^m_n/F^m_n$ (in this notation $m$ is not exponent). Then we cast (\ref{general system-NLS}) into:
\begin{eqnarray}\label{2-bil}
i {\bf D_t} G^m_n \cdot F^m_n&=&G^{m+1}_{n+1} F^{m-1}_{n-1}+G^{m-1}_{n-1} F^{m+1}_{n+1}\nonumber\\
(F^m_n)^2+|G^m_n|^2&=&F^{m+1}_{n+1} F^{m-1}_{n-1},
\end{eqnarray}
where $F_n^m$ is real function, while $G_n^m$ is complex valued function.\\
The 1-soliton solution is:
$$q^m_n=\frac{G^m_n}{F^m_n}=\frac{e^{\eta_1}}{1+\frac{1}{2[\cosh(k_1+p_1+k_1^*+p_1^*)-1]}e^{\eta_1+\eta_1^*}}, \quad \eta_1=k_1 n+p_1 m +\omega_1 t+\eta_1^{(0)},$$
where:
$$ \omega_1=-2 i \cosh(k_1+p_1), \quad (\forall)\quad k_1,  p_1 \in \mathbb{C}.$$\\
For 2-soliton solution we obtain:
\begin{eqnarray*}
G^m_n\!\!\!&\!=\!\!\!\!&e^{\eta_1}+e^{\eta_2}+e^{\eta_1+\eta_2+\eta_2^*+\phi_{12}+\phi_{14}+\phi_{24}} +e^{\eta_1+\eta_2+\eta_1^*+\phi_{12}+\phi_{13}+\phi_{23}}\nonumber\\
F^m_n\!\!\!&\!=\!\!\!\!&1+e^{\eta_1+\eta_1^*+\phi_{13}}+e^{\eta_2+\eta_2^*+\phi_{24}}+e^{\eta_1+\eta_2^*+\phi_{14}}+ e^{\eta_2+\eta_1^*+\phi_{23}}+ e^{\eta_1+\eta_1^*+\eta_2+\eta_2^*+\sum\limits^4_{1\leq i<j}\phi_{ij}} \nonumber
\end{eqnarray*}
where:
$$e^{\phi_{ij}} = \begin{cases} \frac{1}{2} \left[\cosh(k_i+p_i+k_j^*+p_j^*)-1\right]^{-1}, & \mbox{if } i=1,2 \mbox{ and } j=3,4; \vspace{0.3cm}\\ 
2\left[\cosh(k_i+p_i-k_j-p_j)-1\right], & \mbox{if } i=1,2 \mbox{ and } j=1,2\\
& \mbox{or } i=3,4 \mbox{ and } j=3,4; \end{cases}$$
\begin{equation}\label{coeficienti redusi}
\eta_j=k_j n+p_j m+ \omega_j t+\eta_j^{(0)}, \quad j=1,2
\end{equation}
$$ \omega_j=-2 i \cosh(k_j+p_j), \quad k_{j+2}=k_j^*, \quad p_{j+2}=p_j^*.$$
The three-soliton solution is given by:
\begin{eqnarray*}
G^m_n\!\!\!&\!=\!\!\!\!&e^{\eta_1}+e^{\eta_2}+e^{\eta_3}+\nonumber\\
&+& e^{\eta_1+\eta_2+\eta_2^*+\phi_{12}+\phi_{15}+\phi_{25}} +e^{\eta_2+\eta_1+\eta_1^*+\phi_{12}+\phi_{14}+\phi_{24}}+e^{\eta_3+\eta_1+\eta_1^*+\phi_{13}+\phi_{14}+\phi_{34}}+\nonumber\\
&+&e^{\eta_1+\eta_3+\eta_3^*+\phi_{13}+\phi_{16}+\phi_{36}}+e^{\eta_2+\eta_3+\eta_3^*+\phi_{23}+\phi_{26}+\phi_{36}} +e^{\eta_3+\eta_2+\eta_2^*+\phi_{23}+\phi_{25}+\phi_{35}}+\nonumber\\
&+&e^{\eta_1+\eta_2+\eta_3^*+\phi_{12}+\phi_{16}+\phi_{26}} +e^{\eta_1+\eta_3+\eta_2^*+\phi_{13}+\phi_{15}+\phi_{35}}+e^{\eta_2+\eta_3+\eta_1^*+\phi_{23}+\phi_{24}+\phi_{34}}+\nonumber\\
&+&e^{\eta_1+\eta_2+\eta_3+\eta_1^*+\eta_2^*+\phi_{12}+\phi_{13}+\phi_{14}+\phi_{15}+\phi_{23}+\phi_{24}+\phi_{25}+\phi_{34}+\phi_{35}+\phi_{45}}+ \nonumber\\
&+&e^{\eta_1+\eta_2+\eta_3+\eta_1^*+\eta_3^*+\phi_{12}+\phi_{13}+\phi_{14}+\phi_{16}+\phi_{23}+\phi_{24}+\phi_{26}+\phi_{34}+\phi_{36}+\phi_{46}}+ \nonumber\\
&+&e^{\eta_1+\eta_2+\eta_3+\eta_2^*+\eta_3^*+\phi_{12}+\phi_{13}+\phi_{15}+\phi_{16}+\phi_{23}+\phi_{25}+\phi_{26}+\phi_{35}+\phi_{36}+\phi_{56}} \nonumber\\
F^m_n\!\!\!&\!=\!\!\!\!&1+e^{\eta_1+\eta_1^*+\phi_{14}}+e^{\eta_2+\eta_2^*+\phi_{25}}+e^{\eta_3+\eta_3^*+\phi_{36}}+e^{\eta_1+\eta_2^*+\phi_{15}}+ e^{\eta_1+\eta_3^*+\phi_{16}}+ \nonumber\\
&+&e^{\eta_2+\eta_1^*+\phi_{24}}+e^{\eta_2+\eta_3^*+\phi_{26}}+e^{\eta_3+\eta_1^*+\phi_{34}}+e^{\eta_3+\eta_2^*+\phi_{35}}+\nonumber\\
&+&e^{\eta_1+\eta_2+\eta_1^*+\eta_2^*+\phi_{12}+\phi_{14}+\phi_{15}+\phi_{24}+\phi_{25}+\phi_{45}}+e^{\eta_1+\eta_2+\eta_1^*+\eta_3^*+\phi_{12}+\phi_{14}+\phi_{16}+\phi_{24}+\phi_{26}+\phi_{46}}+\nonumber\\
&+&e^{\eta_1+\eta_2+\eta_2^*+\eta_3^*+\phi_{12}+\phi_{15}+\phi_{16}+\phi_{25}+\phi_{26}+\phi_{56}}+e^{\eta_1+\eta_3+\eta_1^*+\eta_2^*+\phi_{13}+\phi_{14}+\phi_{15}+\phi_{34}+\phi_{35}+\phi_{45}}+\nonumber\\
&+&e^{\eta_1+\eta_3+\eta_1^*+\eta_3^*+\phi_{13}+\phi_{14}+\phi_{16}+\phi_{34}+\phi_{36}+\phi_{46}}+e^{\eta_1+\eta_3+\eta_2^*+\eta_3^*+\phi_{13}+\phi_{15}+\phi_{16}+\phi_{35}+\phi_{36}+\phi_{56}}+\nonumber\\
&+&e^{\eta_2+\eta_3+\eta_1^*+\eta_2^*+\phi_{23}+\phi_{24}+\phi_{26}+\phi_{34}+\phi_{35}+\phi_{45}}+e^{\eta_2+\eta_3+\eta_1^*+\eta_3^*+\phi_{23}+\phi_{24}+\phi_{26}+\phi_{34}+\phi_{36}+\phi_{46}}+\nonumber\\
&+&e^{\eta_2+\eta_3+\eta_2^*+\eta_3^*+\phi_{23}+\phi_{25}+\phi_{26}+\phi_{35}+\phi_{36}+\phi_{56}}+e^{\eta_1+\eta_2+\eta_3+\eta_1^*+\eta_2^*+\eta_3^*+\sum\limits^6_{1\leq i<j}\phi_{ij}} \nonumber
\end{eqnarray*}
where:
$$e^{\phi_{ij}} = \begin{cases} \frac{1}{2} \left[\cosh(k_i+p_i+k_j^*+p_j^*)-1\right]^{-1}, & \mbox{if } i=1,2,3 \mbox{ and } j=4,5,6; \vspace{0.3cm}\\ 
2\left[\cosh(k_i+p_i-k_j-p_j)-1\right], & \mbox{if } i=1,2,3 \mbox{ and } j=1,2,3\\
& \mbox{or } i=4,5,6 \mbox{ and } j=4,5,6; \end{cases}$$
\begin{equation}
\eta_j=k_j n+p_j m+ \omega_j t+\eta_j^{(0)}, \quad j=1,2,3
\end{equation}
$$ \omega_j=-2 i \cosh(k_j+p_j), \quad k_{j+3}=k_j^*, \quad p_{j+3}=p_j^*.$$
The $N$-soliton solution has the following form for $G^m_n$ and $F^m_n$ (being essentially the multisoliton solution of Ablowitz-Ladik):
\begin{eqnarray*}
G^m_n&=&\sum\limits_{\mu=0,1}D_2(\underline{\mu})\exp\left({\sum\limits_{i=1}^{2N}\mu_i \eta_i+\sum\limits^{2N}_{1\leq i<j}\mu_i\mu_j\phi_{ij}}\right)\\
F^m_n&=&\sum\limits_{\mu=0,1}D_1(\underline{\mu})\exp\left({\sum\limits_{i=1}^{2N}\mu_i \eta_i+\sum\limits^{2N}_{1\leq i<j}\mu_i\mu_j\phi_{ij}}\right)\\
\end{eqnarray*}
where:
$$e^{\phi_{ij}} = \begin{cases} \frac{1}{2} \left[\cosh(k_i+p_i+k_j^*+p_j^*)-1\right]^{-1}, \\
\hspace{3cm} \mbox{if } i=1,...,N \mbox{ and } j=N+1,...,2N; \vspace{0.3cm}\\ 
2\left[\cosh(k_i+p_i-k_j-p_j)-1\right], \\
\hspace{3cm} \mbox{if } i=1,...,N  \mbox{ and } i=1,...,N \\
\hspace{3cm} \mbox{or } i=N+1,...,2N \mbox{ and } j=N+1,...,2N; \end{cases}$$
\begin{equation}
\eta_j=k_j n+p_j m+ \omega_j t+\eta_j^{(0)}, \quad \eta_{j+N}=\eta_j^*, \quad j=1,...,N
\end{equation}
$$ \omega_j=-2 i \cosh(k_j+p_j), \quad \omega_{j+N}=\omega_j^*,  \quad k_{j+N}=k_j^*, \quad p_{j+N}=p_j^*.$$
and where:
$$D_1(\underline{\mu}) = \begin{cases} 1, & \mbox{when } \sum\limits_{i=1}^N \mu_i=\sum\limits_{i=1}^N \mu_{i+N}; \\
0 & \mbox{otherwise}; \end{cases}$$
$$D_2(\underline{\mu}) = \begin{cases} 1, & \mbox{when } \sum\limits_{i=1}^N \mu_i=1+\sum\limits_{i=1}^N \mu_{i+N}; \\
0 & \mbox{otherwise}. \end{cases}$$

Now, all the results are coming straightforward from the diagonal Ablowitz-Ladik (\ref{general system-NLS}). 
This can be seen immediately looking at the two bilinear systems (\ref{2-bil}) and (\ref{N-bil}). 
The systems are the same (considering the bar to be the up-shift/down-shift in $m$-direction). 
In the case $N=2$, the $m$-dependence is dropped, $p$ from the definition of $\eta$ or $\omega(k,p)$ will be $-i \pi$  or $+i \pi$, 
making the dispersion relation to have two branches (allowing solitons to move either in the same direction or in the opposite one). 
In the case $N=3$, we have similarly $p$, its exponentials are the cubic roots of the unity. 
The Lax pairs can be obtained in the same way but imposing periodicity on the spectral functions  namely:
$$
\left(
\begin{array}{c}
\psi_1(n+1,m+1)\\
\psi_2(n+1,m+1)\\
\end{array}\right)=L_{n,m}\left(
\begin{array}{c}
\psi_1(n,m)\\
\psi_2(n,m)\\
\end{array}\right)
$$
$$
\partial_t\left(
\begin{array}{c}
\psi_1(n,m)\\
\psi_2(n,m)\\
\end{array}\right)=B_{n,m}\left(
\begin{array}{c}
\psi_1(n,m)\\
\psi_2(n,m)\\
\end{array}\right).
$$
Imposing, let us say $N=3$ periodicity, we get $\psi_i(n,m)\equiv u^i_{1}(n), \: \psi_i(n,m+1)=u^i_{2}(n), \:
\psi_i(n,m+2)=u^i_{3}(n), \: \psi_i(n,m+3)=u^i_1(n), \: etc.$ for any $i=1,2$. We will obtain 6 equations and $6\times 6$ Lax pair matrices that we have already found above (\ref{Lax 3 cuplat}).

\section{Time discretization}
We are going to perform fully integrable discretzation of our system ($t$ discretization) using again the Hirota bilinear formalism \cite{hir}. 
It is easy to discretize the bilinear equation. Just replace the continuous bilinear operator with a discrete one and impose the 
gauge-invariance (i.e. invariance with respect to the multiplication with exponential of linears) \cite{hir}:
\begin{equation*}
{\bf D_t} a \cdot b \rightarrow \frac{\tilde a b-a \tilde b}{\delta},
\end{equation*}
where $\delta$ is the discrete step.
\begin{equation*}
{\bf D_t} a \cdot b=a_t b-a b_t=\frac{\tilde a -a}{\delta} b-a \frac{\tilde b -b}{\delta}=\frac{1}{\delta}(\tilde a b-a \tilde b),
\end{equation*}
where $\tilde a=a((m+1)\delta,n)$ and $t$ becomes $m \delta$.

So, our general system (\ref{sys1}), as mentioned in (\ref{N-bil}), has the following bilinear form:
\begin{eqnarray*}
i {\bf D_t} G_\mu \cdot F_\mu&=&\overline{G_{\mu+1}} \underline{F_{\mu-1}} + \underline{G_{\mu-1}} \overline{F_{\mu+1}} \nonumber\\
F_\mu^2+|G_\mu|^2&=&\overline{F_{\mu+1}} \underline{F_{\mu-1}},
\end{eqnarray*}
where $\mu=1,...,N, \: F_0=F_N, \: F_{N+1}=F_1,\: G_0=G_N, \: G_{N+1}=G_1$.

Discretizing the first bilinear equation we get:
\begin{eqnarray*}
i (\widetilde G_\mu F_\mu -G_\mu \widetilde F_\mu)=\delta(\widetilde {\overline{G_{\mu+1}}} \underline{F_{\mu-1}}+\underline{G_{\mu-1}}\widetilde {\overline{F_{\mu+1}}} ) \nonumber,
\end{eqnarray*}
where $F_\mu=F_\mu(m\delta,n)$, $G_\mu=G_\mu(m\delta,n)$, $\widetilde {\overline{G_{\mu}}}=G_\mu((m+1)\delta,n+1)$ and  $\widetilde {\overline{F_\mu}}=F_\mu((m+1)\delta,n+1)$.

Using Hirota-Tsujimoto approach \cite{hir}, \cite{DNLS} we are not changing the second bilinear equation (otherwise the bilinear system will not have two-soliton solution). Our fully discrete bilinear system will be:
\begin{eqnarray}
i (\widetilde G_\mu F_\mu-G_\mu \widetilde F_\mu)&=&\delta(\widetilde {\overline{G_{\mu+1}}} \underline{F_{\mu-1}}+\underline{G_{\mu-1}}\widetilde {\overline{F_{\mu+1}}} ) \label{dis}\\
F_\mu^2+|G_\mu|^2&=&\overline{F_{\mu+1}} \underline{F_{\mu-1}}. \label{con}
\end{eqnarray}

Integrability of the bilinear system (\ref{con}), (\ref{dis}) can be seen from the existence of 3-soliton solutions, which has the same form as (\ref{solutie generalaa}), namely:

\begin{eqnarray}\label{solutie generalaa}
G_{\mu}&=&\sum\limits_{\nu=0,1}D_2(\underline{\nu})\exp\left({\sum\limits_{i=1}^{6}\nu_i [\eta_i+(\mu-1)\log(\epsilon_i)]+\sum\limits^{6}_{1\leq i<j}\nu_i\nu_j\phi_{ij}}\right)\nonumber\\
F_{\mu}&=&\sum\limits_{\nu=0,1}D_1(\underline{\nu})\exp\left({\sum\limits_{i=1}^{6}\nu_i [\eta_i+(\mu-1)\log(\epsilon_i)]+\sum\limits^{6}_{1\leq i<j}\nu_i\nu_j\phi_{ij}}\right)
\end{eqnarray}
where $D_1(\underline{\nu})$, $D_2(\underline{\nu})$, $\eta_i$, $\phi_{ij}$ are given in (\ref{coeficienti generali})
with the difference that now we have $N$ branches of dispersion for each soliton  with the following fully discrete form:
\begin{equation}
\exp{\delta \: \omega_j(k_j)}=\frac{i-\delta \epsilon_j^{-1} e^{-k_j}}{i+\delta \epsilon_j e^{k_j}}, \quad \epsilon_j \in \left\{e^{l \frac{2 \pi i}{N}} \right\},  \: l=1,...,N \: j=1,2,3 \nonumber 
\end{equation}
where  $k_j$ is the wave number and $j$ is the index of the soliton. The Lax pair of this systems is unknown to us. However because we have three soliton solution we expect the system to be completely integrable.

Dividing (\ref{dis}) by $F_\mu \widetilde F_\mu$ and putting $q_\mu=G_\mu/F_\mu$ and $\Gamma_\mu={\widetilde {\overline{F_{\mu+1}}} \underline {F_{\mu-1}} }/{F_\mu \widetilde F_\mu}$, we get:
\begin{eqnarray}
i (\widetilde{q_\mu} -q_\mu)&=&\delta (\widetilde {\overline{q_{\mu+1}}}+\underline{q_{\mu-1}}) \Gamma_\mu \label{comp1}\\
1+|q_\mu|^2&=&\frac{{\overline{F_{\mu+1}}} \underline {F_{\mu-1}} }{F_\mu^2}. \label{comp2}
\end{eqnarray}

But, using (\ref{comp2}) we get:
\begin{equation*}\label{comp3}
\frac{\overline{\Gamma_{\mu+1}}}{\Gamma_\mu}=\frac{\widetilde{\overline{\overline{F_{\mu+2}}}} \: \:F_{\mu}}{\overline{F_{\mu+1}} \:\widetilde{\overline{F_{\mu+1}}} } \:\:  \frac{F_\mu \:\:\widetilde F_\mu}{\widetilde {\overline{F_{\mu+1}}} \: \:\underline{F_{\mu-1}} } 
=\left(  \frac{\widetilde{\overline{\overline{F_{\mu+2}}}} \: \: \widetilde{F_{\mu}}} { \:\widetilde{\overline{F_{\mu+1}}}^2 } \right) \left ( \frac{F_\mu ^2 }{\overline{F_{\mu+1}} \: \: \underline{F_{\mu-1}} }\right)= \frac {1+|\widetilde{\overline{q_{\mu+1}}}|^2}{1+|q_\mu|^2}.
\end{equation*}

Finally, our fully discrete Ablowitz-Ladik is  the following system:
\begin{eqnarray*}
i (\widetilde{q_\mu} -q_\mu)&=&\delta (\widetilde {\overline{q_{\mu+1}}}+\underline{q_{\mu-1}}) \Gamma_\mu\\
\frac{\overline{\Gamma_{\mu+1}}}{\Gamma_\mu}&=& \frac {1+|\widetilde{\overline{q_{\mu+1}}}|^2}{1+|q_\mu|^2}, \quad \mu=1,...,N.
\end{eqnarray*}
Now, eliminating $\Gamma_\mu$, we will obtain the following (higher order) fully discrete Ablowitz-Ladik system:

\begin{equation}
\left(\frac{\widetilde{\overline{q_{\mu+1}}}-\overline{q_{\mu+1}}}{\widetilde{\overline{\overline{q_{\mu+2}}}}+q_\mu}\right)\left(\frac{\widetilde{\overline{q_{\mu+1}}}+\underline{q_{\mu-1}}}{\widetilde{q_{\mu}}-q_\mu}\right) \left(\frac{1+|q_\mu|^2}{1+|\widetilde{\overline{q_{\mu+1}}}|^2} \right)=1, \quad \mu=1,..N.
\end{equation}

\section{Conclusions}
In this paper we studied coupled Ablowitz-Ladik equations with branched dispersion relations. The main motivation was to see that the integrability survives and how to study collisions of solitons in different directions. It was shown by Hirota bilinear formalism and Lax pairs that the system is integrable and moreover it was shown that with a very simple periodic reduction of an 2D integrable Ablowitz-Ladik equation all the results are recovered. This simple method was applied initially in \cite{toki-fane} but we think that it can be widely extended to a lot of semidiscrete and discrete equations to obtain integrable matrix systems with branched dispersions. In fact, for any completely integrable semidiscrete equation
$F(\dot u_n, u_n,u_{n+1},u_{n-1})=0$, its 2D variant $F(\dot u_{n,m} u_{n,m}u_{n+1,m+1},u_{n-1,m-1})=0$, if it is integrable, then by various periodic reduction in $m$, one can obtain integrable matrix systems. 

From the point of view of applications we do not know any physical application. However, because the semidiscrete NLS can model motion of discrete curves \cite{doliwa} we expect that the systems described in this paper can describe motion of 
{\it modular} polymers (containing different types of segments), exactly as in the case of modular gene networks which can be modelled \cite{toki-fane} by semidiscrete KdV systems with branched dispersion. 
We intend to do this in a future publication.

\vskip 0.5cm
{\bf Acknowledgements:} The work is supported by the project PN-II-ID-PCE-2011-3-0083, Romanian Ministery of Education and Research. ASC is supported by the project PN-II-ID-PCE-2011-3-0137, 
Romanian Ministery of Education and Research.
Also we are very grateful to the referees for all the obervations.

\end{document}